\documentclass{aa} 

\usepackage{graphicx} 
\usepackage[varg]{txfonts}
\usepackage{natbib}
\begin{document} %

   \title{Millimeter radiation from a 3D model of the solar atmosphere}

   \subtitle{II. Chromospheric magnetic field}

   \author{M. Loukitcheva
          \inst{1,2,3}
         \and
          S. M. White \inst{4}
           \and
          S. K. Solanki \inst{1,5}
          \and
          G. D. Fleishman \inst{2}
          \and
           M. Carlsson  \inst{6}
            }

   \institute{Max-Planck-Institut f\"ur Sonnensystemforschung, Justus-von-Liebig-Weg 3,
   37077 G\"ottingen, Germany\\
    \email{lukicheva@mps.mpg.de}
    \and
    Center For Solar-Terrestrial Research, New Jersey Institute of Technology,
    323 Martin Luther King Boulevard, Newark, NJ 07102, United States
    \and
    Astronomical Institute, St.Petersburg University, Universitetskii pr. 28, \\ 198504 St.Petersburg, Russia
    \and
    Space Vehicles Directorate, Air Force Research Laboratory, Kirtland AFB, NM, United States
    \and
    School of Space Research, Kyung Hee University, Yongin, Gyeonggi 446-701, Korea
    \and
    Institute of Theoretical Astrophysics, University of Oslo, P.O. Box 1029, Blindern, N-0315 Oslo, Norway
             }

\titlerunning{Chromospheric magnetic field from a three-dimensional model}
\authorrunning{M. Loukitcheva et al.}

\date{Received 11 June 2016 / Accepted 8 February 2017}


  \abstract
   {}
   {We use state-of-the-art, three-dimensional non-local thermodynamic equilibrium (non-LTE) radiative magnetohydrodynamic simulations of the quiet solar atmosphere to carry out detailed tests of chromospheric magnetic field diagnostics from free-free radiation at millimeter and submillimeter wavelengths (mm/submm).}
   { The vertical component of the magnetic field was deduced from the mm/submm brightness spectra and the degree of circular polarization synthesized at millimeter frequencies. We used the frequency bands observed by the Atacama Large Millimeter/Submillimeter Array (ALMA) as a convenient reference. The magnetic field maps obtained describe the longitudinal magnetic field at the effective formation heights of the relevant wavelengths in the solar chromosphere.}
   { The comparison of the deduced and model chromospheric magnetic fields at the spatial resolution of both the model and current observations demonstrates a good correlation, but has a tendency to underestimate the model field. The systematic discrepancy of about 10\% is probably due to averaging of the restored field over the heights contributing to the radiation, weighted by the strength of the contribution. On the whole, the method of probing the longitudinal component of the magnetic field with free-free emission at mm/submm wavelengths is found to be applicable to measurements of the weak quiet-Sun magnetic fields. However, successful exploitation of this technique requires very accurate measurements of the polarization properties (primary beam and receiver polarization response) of the antennas, which will be the principal factor that determines the level to which chromospheric magnetic fields can be measured. }
   {Consequently, high--resolution and high--precision observations of circularly polarized radiation at millimeter wavelengths can be a powerful tool for producing chromospheric longitudinal magnetograms. }


   \maketitle

\section{Introduction}

Images of the solar chromosphere in traditional optical and ultraviolet lines vividly demonstrate a clear organization of the chromospheric plasma along magnetic structures \citep[see, e.g.,][]{Rutten2007,depontieu2007}. Both observations and simulations show that the chromospheric magnetic field plays a fundamentally important role in chromospheric heating, chromospheric dynamics, and the propagation of waves through the chromosphere into the corona \citep[e.g., see][for a review]{wiegelmann}. However, mainly owing to the shortcomings of the currently available diagnostics, the chromospheric magnetic field is still poorly known.

Optical polarimeters making use of Zeeman splitting of spectral lines provide a reliable source of information on photospheric magnetic field. The situation is less ideal in the chromosphere. The few Zeeman-sensitive chromospheric lines, for example, Ca II 854.2 nm and the He I 1083~nm triplet, which can be employed for chromospheric magnetometry, are commonly characterized by a weak polarization signal, often restricting their use in the quiet Sun (QS). In addition, many of these lines suffer from a wide range of formation heights and require a careful treatment of non-local thermodynamic equilibrium (non-LTE) effects. Most recent works on the chromospheric magnetic field focus on the members of the Ca II IR triplet lines \citep[849.8 nm, 854.2 nm, and 866.2 nm, e.g.,][]{socas,carlin} and on the He I 1083 nm line \citep[e.g.,][]{solanki,Martinez,Schad}.

A complementary method of measuring magnetic fields in the solar chromosphere is based on the analysis of bremsstrahlung (free-free) radio emission, which becomes polarized in the presence of a  magnetic field. Observations of free-free polarization effects at mm/submm wavelengths carry information about the longitudinal component of the magnetic field at the chromospheric heights at which the radiation is formed. The method was developed by \citet{bogod80} and was initially tested using observations of plages with the radiotelescope RATAN- 600 in the wavelength range of $2-4$~cm at a spatial resolution in the range $17-34$\arcsec\ (limited to one spatial dimension). \citet{bogod80} measured a magnetic field of about 50~G in the transition region of plages, with a standard deviation of 10~G. This magnetic field was found to be consistent with the Mt.Wilson photospheric magnetic field measurements within a factor of two, although the measured photospheric field is weaker than typical plage magnetic field  \citep[see, e.g.,][]{buehler}. \citet{Gelfreikh} provided further examples of the effectiveness of the method, which included observations of chromospheric and coronal structures such as prominences, coronal holes, CMEs, loops, and arches with RATAN- 600. The intensity and circular polarization maps obtained with the Nobeyama Radio Heliograph (NoRH) at 17~GHz (1.76~cm) at a resolution of about 15\arcsec\ were used for magnetography of transition region structures such as isolated spots, plages, and bipolar active regions \citep{grebinskij1998, gelfreikh1999,Iwai}. For short cm wavelengths \citet{grebinskij2000} developed a technique to distinguish between contributions of the opaque chromosphere and the optically thin corona in the observed polarized emission, if spectral intensity and polarization measurements are available. This technique also delivered practical estimates of magnetic fields in the absence of spectral information. 
In general, extensive earlier studies have demonstrated that the precise measurement of the brightness spectrum and the degree of polarization are jointly needed to measure the longitudinal component of the magnetic field at the height where emission is generated.

 At cm wavelengths gyroresonance emission at the second or third harmonics of the electron gyrofrequency can also contribute significantly above sunspots and thus can mask the free-free contribution. This complicates the use of cm wavelengths for chromospheric magnetic field measurements, and the relative contributions of the two mechanisms are widely discussed in the literature \citep[e.g.,][]{aliss,Schonfeld}.
  Above  sunspots gyroresonance emission cannot be neglected for wavelengths $\lambda \geq 2$~cm, while for the weaker magnetic fields in plage and quiet regions it becomes important only starting from $\lambda \geq 5$~cm \citep{Gelfreikh}. The radiation of the non-flaring Sun at mm/submm wavelengths comes primarily from the chromosphere with its gentle temperature gradient and is purely due to the bremsstrahlung mechanism.

Up to now the technique of deriving longitudinal magnetic fields from bremsstrahlung was used solely for observational data at cm wavelengths with the limitations discussed above. The critical issue for mm/submm wavelengths is the relatively low degree of polarization produced by free-free radiation and the need for accurate brightness temperature spectra. The Atacama Large Millimeter/Submillimeter Array (ALMA) operating at these wavelengths will be able to produce high quality brightness temperature spectra at much better spatial resolution than is available in any prior observations. In this paper we use models to predict the polarization that can be observed at millimeter wavelengths. The models are advanced three-dimensional (3D) radiative magnetohydrodynamic (MHD) simulations of the enhanced network made with the Bifrost code \citep{gudiksen,carlsson16}.  We calculate the mm/submm emission produced by the free-free mechanism from the model physical parameters and then analyze the resulting images as if they were observed data to deduce the longitudinal magnetic field. The deduced magnetic field parameters are then compared directly with the model field to assess the precision of the method.  At the time of writing the ALMA project has not finalized commissioning of circular polarization measurements and so the practical capabilities for such observations are not yet available.  The conclusions presented in this paper originate purely from the results of simulations. For practical estimations regarding potential use of ALMA for magnetic field measurements, comparison of the simulated polarization with realistic ALMA polarization measurements and their uncertainties must be carried out when the latter are available.

This is the second paper in the series exploring the diagnostic potential of mm/submm observations for studies of the solar chromosphere.
In Paper I \citep{LSC15}, we carried out detailed tests of chromospheric thermal structure diagnostics at mm/submm wavelengths using simulations of the solar atmosphere performed with the Bifrost code. We found that the millimeter brightness temperature is a reliable measure of the gas temperature at the effective formation height at a given location on the solar surface when observed at a spatial resolution of 1\arcsec\ or better. In this paper, we focus on diagnostics of magnetic fields at the heights where mm/submm radiation is formed in the context of requirements for successful chromospheric magnetic field measurements. We also reconsider the assumptions made in Paper I for the computation of the mm/submm radiation, in particular, the use of the quasilongitudinal (QL) approximation and unity refractive index.

The structure of the paper is as follows. In Sects.~2 and 3 we describe the 3D radiative MHD model atmosphere based on the Bifrost code and the mm/submm polarized radiative transfer computations. In Sect.~4 we present the method for recovering the vertical component of the magnetic field from the observed free-free radiation. The recovered magnetic field, the accuracy of the restoration, and the effect of spatial smearing of the model polarization are discussed in Sect.~\ref{res}.  In Sect.~\ref{alma} we describe in detail how ALMA and similar interferometric radio telescopes measure circular polarization, and we discuss the potential limitations of chromospheric magnetic field measurements with ALMA. We summarize our findings and present conclusions about the prospects of probing chromospheric longitudinal magnetic field with the free-free emission at mm/submm wavelengths in Sect.~\ref{discuss}.

\section{Three-dimensional model atmosphere}

We study the formation of mm/submm continuum in the magnetized solar atmosphere using snapshot 385 of the simulation \emph{en024048{\_}hion} performed with the Bifrost code \citep{gudiksen} and made available through the Interface Region Imaging Spectrograph \citep[IRIS;][]{De Pontie} project at the Hinode Science Data Centre Europe (http://sdc.uio.no). The same snapshot was used in Paper I and in a series of papers investigating the formation of IRIS diagnostics \citep{leenaarts13a,leenaarts13b,pereira,pereira15,Rathore,Rathore1,Lin,Rathore2} and the formation of other chromospheric lines \citep{leenaarts12,leenaarts15,rodriguez,stepan,stepan2015}. This snapshot has been described in detail by \citet{carlsson16} and in Paper I, so here we restrict ourselves to a brief summary.

The simulation covers a physical extent of 24~x~24~x~16.8 Mm with a grid of 504~x~504~x~496 cells, extending from 2.4~Mm below the photosphere (by convention at height $Z=0$) to 14.4~Mm above, so that it covers the upper convection zone, photosphere, chromosphere, and the lower corona. The horizontal axes have an equidistant grid spacing of 48~km (0.06\arcsec), while the vertical grid spacing is non-uniform with a spacing of 19~km between $Z = -1$ and $Z = 5$~Mm. The spacing increases toward the bottom and top of the computational domain to a maximum of 98 km. The simulation contains a magnetic field with an average unsigned strength of 50~G in the photosphere that is concentrated in the photosphere in two clusters of opposite polarity lying 8~Mm apart, representing two patches of QS network. In the following we refer to the analyzed snapshot from this simulation simply as the 3D MHD snapshot.

\section{Synthesis of mm brightness}\label{sect_GEN}

   Radiative transfer through the simulations is necessary to determine the expected emission characteristics at mm/submm wavelengths. The emission was calculated under the assumption that in the quiet Sun
bremsstrahlung (free-free) opacity is responsible for the mm/submm continuum radiation.
We distinguish two sources of opacity: due to encounters between free electrons and singly ionized ions ($ei$-opacity) and between free electrons and neutral hydrogen (H$^-$ opacity).
The corresponding absorption coefficients $\chi$ depend on the effective
number of collisions undergone by an electron per unit time, including collisions with ions ($\chi_\mathrm{ep}\propto \frac{N_\mathrm{e}^2}
{T_\mathrm{e}^{3/2} \nu^{2}}$) and with neutral hydrogen atoms ($\chi_\mathrm{eH}\propto \frac{N_\mathrm{e}
\,N_\mathrm{H}\,T_\mathrm{e}^{1/2}}{\nu^{2}}$), where $T_\mathrm{e}$ is the electron temperature, $N_\mathrm{e}$ and $N_\mathrm{H}$ are the concentrations of electrons and hydrogen atoms, respectively, and $\nu$ is the observing frequency.

In magnetized plasma, radiation propagates via two oppositely polarized modes of electromagnetic wave: the extraordinary ($x$-mode) and the ordinary ($o$-mode), usually referred to as natural or normal modes \citep[cf.][]{zheleznyakov}.
Free-free emission in the presence of magnetic fields becomes polarized with different opacities for the $x$- and the $o$-mode.\footnote{At the location where the emission is formed the $x$- and $o$- eigen-modes are in general elliptically polarized. Measurements of the polarization ellipse can provide information about the transverse component of the magnetic field. However, while propagating out through the non-uniform solar atmosphere, any polarization ellipse evolves toward a circle, which is a 'limiting
polarization' state \citep{zheleznyakov,fleishman}, providing information only about the line-of-sight component of the magnetic field.}.
Instead of the quasilongitudinal approximation used in Paper I, we calculate the absorption coefficients using the generalized formulas with the full anisotropic term included, following \citet{zlotnik}, \citet{zheleznyakov} and \citet{fleishman},
\begin{equation}
\chi_\mathrm{\sigma}=\chi_\mathrm{\sigma}^{0} \frac{F_\mathrm{\sigma}}{n_\mathrm{\sigma}},
 \label{eq_1}
\end{equation}
where subscript $\sigma$ denotes one of the modes ($x$ or $o$), $\chi_\mathrm{\sigma}^{0}$ is the absorption coefficient for the isotropic plasma, $\frac{F_\mathrm{\sigma}}{n_\mathrm{\sigma}}$ is the anisotropic term with $n_\mathrm{\sigma}$ being the refraction index defined as\begin{equation}
n_\mathrm{\sigma}^2=1-\frac{2v(1-v)}{2(1-v)-u\sin^2\theta +\sigma\sqrt{D}},
 \label{eq_2}
\end{equation}
\begin{equation}
 D=u^2\sin^4\theta+4u(1-v)^2\cos^2\theta,
\end{equation}
\begin{equation}
u=\left(\frac{\nu_B}{\nu}\right)^2, v=\left(\frac{\nu_p}{\nu}\right)^2.
\end{equation}
Here $\nu_B$ is the gyrofrequency, $\nu_p=e\sqrt{N_e/(\pi m_e)}$ is the electron plasma frequency, $\sigma=-1$ for the $x$-mode and $\sigma=1$ for the $o$-mode, and $\theta$ is the angle between magnetic field and the line of sight.

The factor $F_\mathrm{\sigma}$ \citep[cf. Eq. 5 from][]{wang} is responsible for the effect of the magnetic field and has the following form:
\begin{equation}
F_\mathrm{\sigma}=2 \frac{\sigma \sqrt{D} [u\sin^2\theta + 2(1-v)^2] -u^2 \sin^4 \theta}{\sigma \sqrt{D}[2(1-v)-u \sin^2 \theta + \sigma\sqrt{D}]^2}
.\end{equation}

For comparison, in the quasilongitudinal approximation, which is valid for all angles between the magnetic field and line of sight except a narrow interval of angles close to transverse propagation \citep{zheleznyakov} the absorption coefficients for $x$- and $o$-modes are determined by the much simpler relation \citep{zlotnik}
  \begin{equation}
   \chi_\mathrm{\sigma} \simeq \frac{\chi_\mathrm{\sigma}^{0}}{(1+\sigma\frac{\nu_B}{\nu}|\cos\theta|)^2}
     \label{eq_QL}
   .\end{equation}

The difference introduced by use of the more exact expression for the mode properties is discussed below in Sec.~\ref{res}.

The radiative transfer in magnetized solar media is considered separately for the $x$-mode and the $o$-mode in terms of the brightness temperature $T_\mathrm{b}$,
  \begin{equation}
   \frac{dT_\mathrm{b}^{\sigma}}{d\tau} =
    T_\mathrm{e} - T_\mathrm{b}^{\sigma}
              ,   \label{eq2}
   \end{equation}
   where $T_\mathrm{e}=T_\mathrm{e}(l)$ is the profile of the kinetic temperature
   along the light path, $\tau(l) = \int^l_{l_\mathrm{0}} \chi_\mathrm{\sigma}(l)dl$~
   is the optical depth, and $l$ is geometrical
  distance along the light path.

For the chromosphere, which is optically thick at mm/submm wavelengths, the brightness of the natural modes can be approximated as follows:
   \begin{equation}
     T_\mathrm{b}^{\sigma}(\nu)=T_\mathrm{e}  \\ \tau(\nu)\geq 1 \label{eqTbT}
   .\end{equation}

   The exact solution of the radiative transfer equation on a discrete grid can be
   expressed in terms of $T_\mathrm{e}$ \citep{hagen} as
   \begin{equation}
   T_\mathrm{b}^{\sigma}=\sum_{r=1}^n(1-\exp^{-\chi_\mathrm{\sigma}^{r}\Delta
   h})\,T_\mathrm{e}^r \exp^{-\sum_{s=1}^{r-1} \chi_\mathrm{s}\Delta h}   ,      \label{eq6}
   \end{equation}
   where $\chi_\mathrm{\sigma}^{r}$ is taken at position $r$, $r=1$ corresponds to the layer  at the top of the
   atmosphere, and $\Delta h$ is grid spacing.
   The items within the sum represent the contribution of various layers to the emergent intensity of the radiation. We
   refer to these items as the values of the (unnormalized) contribution function (CF) to $T_\mathrm{b}^{\sigma}$.

The total brightness temperature $T_\mathrm{b}$ and the degree of polarization $P$
   are defined as
   \begin{equation}
    T_\mathrm{b}=\frac{T_\mathrm{b}^{x}+T_\mathrm{b}^{o}}{2},
    \end{equation}
    \begin{equation}
   P=\frac{T_\mathrm{b}^{x}-T_\mathrm{b}^{o}}{T_\mathrm{b}^{x}+T_\mathrm{b}^{o}}
   \label{eq_p}
   .\end{equation}

   The total brightness temperature $T_\mathrm{b}$ is the mean value of the brightness temperatures in the two natural orthogonal polarizations, which is almost exactly the brightness of plasma with the same distributions of $N_\mathrm{e}$ and $T_\mathrm{e}$ but in the absence of a magnetic field.
   It is practical to rewrite Eq.~\ref{eq_p} in terms of the two orthogonal modes of circular
polarization of measured radiation to explicitly take into account the sign of the polarization corresponding to the sign of the magnetic field,
   \begin{equation}
   P=\frac{T_\mathrm{b}^{R}-T_\mathrm{b}^{L}}{T_\mathrm{b}^{R}+T_\mathrm{b}^{L}},
   \end{equation}
   where $R,L$ denote right circular polarization and left circular polarization, respectively.

We have calculated the polarized emission from the 3D MHD snapshot at frequencies from 638~GHz down to 35~GHz (wavelengths from 0.47~mm to 8.6~mm). We selected 16 frequencies, which correspond to the fixed central frequencies for four spectral windows in each of ALMA Bands 3, 6, 7, and 9 (taken from the ALMA technical Handbook for the Cycle 3 and ALMA Cycle 3 Proposers Guide and Capabilities). The frequency list was filled out with eight (lower) frequencies from ALMA Bands 1 and 2, which have not yet been delivered, but for which work is in progress.
In the calculations we assume each vertical column in the 3D snapshot to be an independent 1D plane-parallel atmosphere.

\section{Method of measuring magnetic field from thermal bremsstrahlung}

Because of the dependence on the magnetic field the free-free absorption coefficient and the corresponding opacity is higher in the $x$-mode than in the $o$-mode (Eq.~\ref{eq_QL}). Consequently, the two modes become optically thick at slightly different heights in the atmosphere. In case of an isothermal plasma this does not matter since both modes still have the same brightness temperature (the temperature of the plasma) and no net polarization results. However, in a plasma with a temperature gradient along the line of sight (LOS) the presence of a magnetic field shifts the effective emitting heights (where $\tau_\mathrm{\sigma}(\nu)\simeq1$) for the two modes in different directions up and down the temperature gradient. The $x$-mode becomes optically thick in higher (hotter) levels of the solar atmosphere with positive temperature gradient, while the $o$-mode comes from lower (cooler) levels. This difference in the opacities, which depends on the strength of the magnetic field, manifests itself as circular polarization. However, the value of the measured polarization reflects the difference in the temperatures between the effective emitting levels of the two modes and itself carries no direct information about the field strength. \citet{bogod80} demonstrated that the temperature gradient can be obtained from the brightness temperature variation with frequency, so that the effect of the magnetic field can be isolated. The proposed method is based on the use of the scaling law, which relates the brightness temperatures of the natural modes $T_\mathrm{b}^{\sigma}$ in a magnetic field to the total unpolarized brightness temperature $T_\mathrm{b}$,\begin{equation}\label{scale}
T_\mathrm{b}^{\sigma}(\nu)=T_\mathrm{b}(\nu+\sigma\nu_B|\cos\theta|).
\end{equation}

Using the scaling law from Eq.~\ref{scale}~ \citet{bogod80} obtained the expression for circular polarization degree $P$ through the logarithmic spectral index $n$ (i.e., the slope of the brightness spectrum),
\begin{equation}\label{P_n}
    P=n\frac{\nu_B}{\nu}\cos\theta,
\end{equation}
where the logarithmic spectral index $n$ is introduced as
\begin{equation}\label{n}
    n\equiv-\frac{\partial(\ln T_\mathrm{b})}{\partial(\ln \nu)}=\frac{\partial(\ln T_\mathrm{b})}{\partial(\ln \lambda)}.
\end{equation}

Here the frequency serves as a proxy for height due to the frequency-dependence of the level of the optically thick layer. Recalling that the electron gyrofrequency is $\nu_B=2.8\times10^6 B$, and that $B_l=B\cos\theta$ is the longitudinal component of the magnetic field,
we can obtain an estimate for the latter from Eq.~\ref{n} as follows:
\begin{equation}\label{Bl}
    B_l=\frac{P \nu}{2.8\times 10^6 n}
\end{equation}

To exploit this technique one needs to measure the degree of polarization at a given frequency and the local slope of the brightness temperature spectrum around this frequency.


It should be noted that the method assumes a homogeneous magnetic field within the layer where the natural modes become optically thick. The width of this layer depends on the local magnetic field strength and is small enough (a few kilometers) to make this approximation reasonable (see Sect.~\ref{res}). The estimated magnetic field refers to the effective emitting height at a given frequency so that the magnetic field at different heights can be measured by changing the frequency. If the degree of polarization $P$ and spectral index $n$ are measured with sufficient precision at a number of frequencies, Eq.~\ref{Bl} provides a method for radio magnetography at chromospheric heights.

At mm wavelengths the method is expected to work effectively both for weak fields (the QS, coronal holes, prominences, and plage areas) and for strong fields in sunspots provided that the recorded polarization is due to the free-free mechanism. However, polarization effects for free-free emission at these high frequencies are weak and the circular polarization degree ranges between 0.1\% and 10\%, both for weak and strong fields \citep[see, e.g.,][]{grebinskij2000}. Hence, the noise level of observations is a major concern for the application of this method to real observational data. For instance, the sensitivity of default 10-sec integrated NoRH images at 17~GHz (17.6~mm) is about 1\%, which leads to an accuracy of the restored magnetograms of about 100~G. However, the sensitivity of NoRH polarization measurements can be substantially improved by averaging the images longer (e.g., for 10~min instead of 10~s) and the potential sensitivity of the method exceeds 1~G \citep{Gelfreikh}. The magnetic fields of a few Gauss in the corona estimated from this method have already been reported using the RATAN-600 spectral-polarization observations \citep{grebinskij2000}. Further discussion of the requirements for the sensitivity of polarization measurements is given in Sects.~\ref{res} and \ref{discuss}.

\section{Results}\label{res}

\begin{figure}
  \centering
            \includegraphics[width=0.4\textwidth,angle=90]{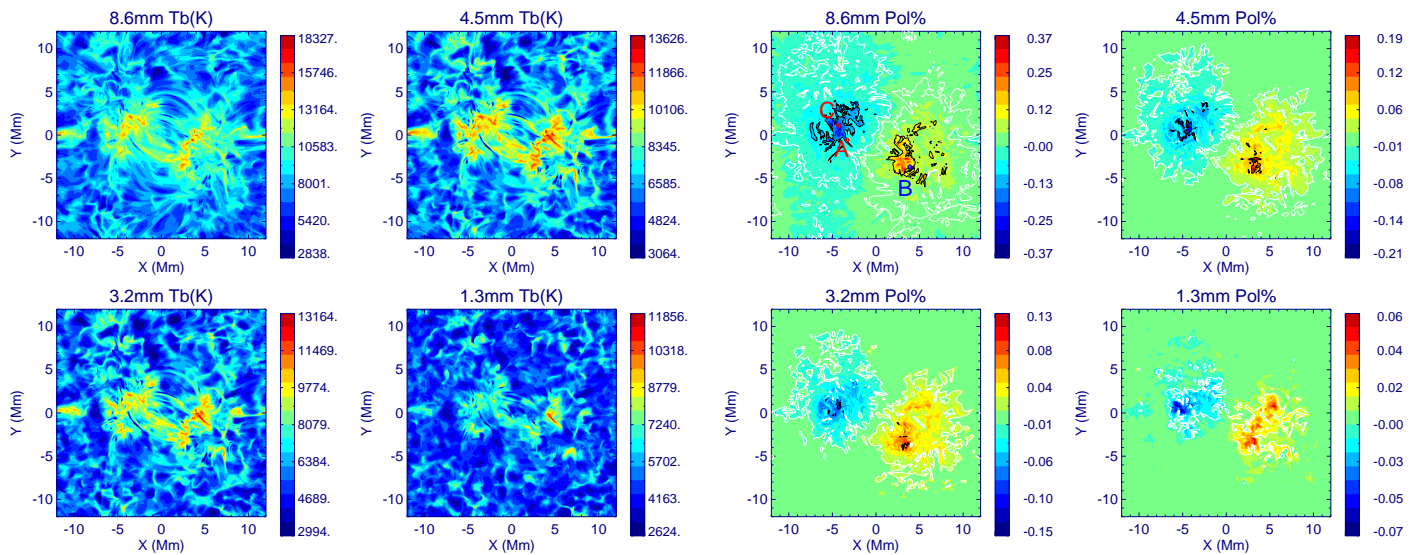}
      \caption{Maps of simulated total brightness at the resolution of the model for the wavelengths 8.6, 4.5, 3.2, and 1.3~mm.
              }
         \label{fig1}
   \end{figure}

In Table~\ref{table:1} we list the average brightness temperature $\langle T_b \rangle$, RMS variation $T_b^{rms}$, and relative brightness temperature contrast $\frac{T_b^{rms}}{<T_b>}$ for the chosen wavelengths derived from the 3D MHD snapshot across the field of view seen in Fig.~\ref{fig1}. The frequencies are not identical to those used in paper I but are sufficiently close that  the results of brightness calculations using the formulae from Sect.~\ref{sect_GEN} can be directly compared with the results presented in Table 1 of Paper I, which were obtained using the QL approximation. The difference between these two approaches is in the treatment of the dependence of the absorption coefficient on the magnetic field and the refractive index, which was assumed to be unity in Paper I. Both approaches produce very similar brightness temperatures for all wavelengths. The relative intensity contrast listed in Table~\ref{table:1} is exactly the same as reported in Paper I. Further we analyzed the relative error in brightness temperature, which was introduced as $dT_b=\frac{T_b^{QL}-T_b^{nQL}}{T_b^{nQL}}$, where $T_b^{QL}$ and $T_b^{nQL}$ are the simulated temperatures from the QL approximation and the exact mode treatment, respectively. The maximum values of the normalized brightness temperature difference are less than 1\% for all wavelengths, varying from 0.5\% at 1~mm up to 0.8\% at 10~mm. Furthermore, we did not find any evidence for larger brightness differences at large angles between the magnetic field and LOS. For the analysis we determined the inclination angles at the effective formation heights for each line of sight and considered only those exceeding 89 deg with respect to the positive z-axis (corresponding to $2-3$\% of the total number of points depending on the wavelength). Thus we confirm the relevance of the QL approximation for the calculation of the free-free absorption coefficient for all angles between the magnetic field and the line of sight, including the angles close to transverse propagation, which is expected given that $\nu_B/\nu \lesssim 0.001 \ll 1$.

The last column of Table~\ref{table:1} lists effective formation heights $<h_{eff}>$ of emission at the  wavelengths investigated, averaged over all spatial locations and over the two natural modes. Effective formation heights for each mode were derived from individual CFs to intensity as the heights of the CF centroids (see Paper I for details). In the 3D MHD snapshot, the maximum difference between the individual $x$- and $o$-mode effective formation heights does not exceed $4$~km at the longest wavelength of 8.6~mm and is only $2$~km at $\lambda\,=\,$3~mm. This height difference determines the width of the layer to which the measured magnetic field refers. The method used here assumes that the magnetic field does not vary significantly across a layer of this thickness, which is consistent with the observation that the gradients in B in the simulation have much larger length scales. Since the derived widths are small, the approximation of homogeneous magnetic field should be reasonable.

Table~\ref{table:11} lists maximum polarization degree $|P_{max}|$ (in \%), percentage of points with $|P|\geq0.01$\% and with $|P|\geq0.1$\%, and maximum brightness temperature difference between the two polarizations $|T_b^R-T_b^L|$ in Kelvin and in \% of the RMS variation $T_b^{rms}$  for the wavelengths in the ALMA bands.

\begin{table}
\caption{Average total brightness temperature $<T_b>$, RMS variation $T_b^{rms}$, relative brightness temperature contrast $\frac{T_b^{rms}}{<T_b>}$, and average effective formation height $<h_{eff}>$ for a number of analyzed wavelengths.}             
\label{table:1}      
\centering                          
\begin{tabular}{l c c c c}        
\hline\hline                 
$\lambda$, mm & $<T_b>$, K & $T_b^{rms}$, K & $\frac{T_b^{rms}}{<T_b>}$  & $<h_{eff}>$, km\\    
\hline                        
\\
\noalign{\smallskip}

0.47 & 4517   &   538  &   0.12  &   750\\
1.3 & 4997   &   957  &   0.19  &    1150\\
3.2 & 6205   &   1360 &   0.22  &    1620\\
4.5 & 6775   &   1483  &   0.22  &   1770\\
8.6 & 8194  &   1729  &   0.21  &   2050\\

\hline                                   
\end{tabular}
\end{table}

\begin{table*}
\caption{Maximum polarization degree (in \%), percentage of points with $|P|\geq0.01$\% and with $|P|\geq0.1$\%, maximum absolute brightness difference between the two polarizations $|T_b^R-T_b^L|$ in Kelvin and in \% of the RMS variation $T_b^{rms}$ for a number of analyzed wavelengths.}             
\label{table:11}      
\centering                          
\begin{tabular}{l c c c c c}        
\hline\hline                 
$\lambda$, mm & $|P_{max}|$, \% & $|P|\geq0.01$\%, \% & $|P|\geq0.1$\%, \% & $|T_b^R-T_b^L|_{max}$, K & $|T_b^R-T_b^L|_{max}$, \% \\    
\hline                        
\\
\noalign{\smallskip}

0.47  & 0.02  & 0.3 & 0 & 3 & 0.6 \\
1.3  & 0.07  & 9   & 0 & 10 & 1.0 \\
3.2  & 0.15  & 27  & 0.1 & 21 & 1.5 \\
4.5  & 0.21 & 35  & 0.5 & 33 & 2.2 \\
8.6  & 0.37 & 51  & 4  & 80 & 4.6 \\

\hline                                   
\end{tabular}
\end{table*}

\subsection{ALMA measurements of polarization}\label{alma}

Before proceeding, it is important to recognize how ALMA and similar interferometric radio telescopes measure circular polarization. In general, ALMA solar observations will have two components: high spatial resolution imaging of relatively small fields of view with the multielement interferometer and relatively low spatial resolution single-dish mapping of the solar disk. The interferometer resolves out large spatial scales and is therefore insensitive to the average emission level of the solar disk so that the denominator in Eq.~\ref{eq_p} is reduced while the numerator is largely unchanged. That is, the interferometer will measure brightness temperature contrast across the solar atmosphere, not absolute brightness temperatures.

The ALMA receivers use linearly polarized feeds, which is an advantage for the precise measurement of circular polarization. In the case of circularly polarized feeds, small differences in gain between the two polarizations are difficult to measure and so errors in amplitude gain determination couple directly into errors in the degree of circular polarization. In the case of linearly polarized (labeled X and Y) feeds, circular polarization is determined from the complex cross-correlations, XY and YX, so amplitude gain differences between X and Y do not affect the result. In the case of the Sun, we have the additional advantage that Faraday rotation is known to remove all linear polarization and thus we can enforce XX=YY. The instrumental polarization-leakage (of linear into circular) terms whose determination is required for precision circular polarimetry are generally stable with time and so only need to be remeasured very occasionally. \citet{Rayner2000} discuss these issues in detail and show that degrees of circular polarization as low as 0.01\% can be measured in practice with linearly polarized feeds.

The degrees of polarization reported in Table~\ref{table:11} are calculated on the absolute temperature scale that includes the (large spatial scale) disk component and are therefore not representative of the degree of polarization that the interferometer will measure. The values that the interferometer will see are closer to the $|T_b^R-T_b^L|_{max}$ values (given in Table~\ref{table:11}) divided by the $T_b^{rms}$ values in Table~\ref{table:1}, i.e., typically a factor of 10 larger than the degrees of polarization on the absolute temperature scale (see $|P_{max}|$ and $|T_b^R-T_b^L|_{max}$ in \% in Table~\ref{table:11}).

However, the interpretation of these observations in terms of magnetic fields will be complicated by several other factors. The high-precision circular polarization measurements reported by \citet{Rayner2000} were for compact sources close to the center of the field of view of the telescope, whereas in the case of the Sun we are generally interested in polarization across the entire field of view. The two linear polarizations might not have identical off-axis responses and, in particular, the ALMA receivers are known to suffer from beam squint, in which  the centers of the primary beams of the two polarizations are slightly offset from each other on the sky. Effects such as this mean that the fidelity of the circular polarization measurements will vary across the primary beam: correcting for this effect will require careful measurement of the beam patterns of the two polarizations independently. Many ALMA solar observations, particularly at the higher frequencies, will use mosaicking techniques to map fields of view larger than a single primary beam, and precise measurements of polarization in such large fields where data from multiple pointings are combined will require excellent knowledge of the beam responses of both polarizations and the use of these responses during the imaging stages. ALMA plans to make such measurements using a polarized beacon located on a mountain peak visible to the array.

While the absence of the solar disk component from the interferometer data provides larger degrees of polarization and therefore better measurements of circular polarization, it also complicates the conversion of those measurements into degrees of polarization on the absolute scale. The determination of magnetic field strength relies on measuring the gradient of brightness temperature with frequency and, hence, the brightness temperature at each frequency must be on the same relative scale. This requires determining the disk level omitted from the interferometer data at each frequency used. The ALMA single-dish measurements will be used for this purpose. While the formal noise level in the single-dish data is less than 1 K (12 m aperture, 2 GHz bandwidth, 1 second integration,  and 7000 K system temperature) and therefore acceptable at the 0.01\% level, in practice an individual single-dish map has a much larger uncertainty owing to the calibration procedure and atmospheric fluctuations; the overall map level will likely be scaled to frequency--dependent values determined from a large number of measurements \citep{White}. In terms of polarization, the resolution of the single-dish maps is much poorer than the interferometer data and therefore averages over much larger areas. As long as the large-scale emission resolved out by the interferometer is unpolarized on average, the background circular polarization level will be zero; we benefit from the fact that the ALMA feeds are linearly polarized since we could not distinguish net background polarization from small gain errors for circularly polarized feeds. The only circumstance in which the large-scale emission resolved out by the interferometer is likely to be circularly polarized is in a large sunspot, particularly at the higher frequency bands with smaller fields of view; in other unipolar regions, such as coronal holes, the individual circularly polarized features are likely to be small, similar to the photospheric magnetic network, and therefore correctly measured by the linearly polarized feeds of the interferometer. In such cases, single-dish data that resolve the relevant large-scale feature may be able to measure the missing polarization level. The single-dish data will be acquired in the on--the--fly mode \citep{White} in which the telescope is continuously moving \citep{Mangum2007}, so the primary beam responses of the polarizations are effectively spatially averaged in the resulting data. Modeling will be needed to assess the impact of this process on circular polarization measurements once the polarization primary beam responses and the polarization leakage terms have been measured.

 \begin{figure}
  \centering
            \includegraphics[width=0.4\textwidth,angle=90]{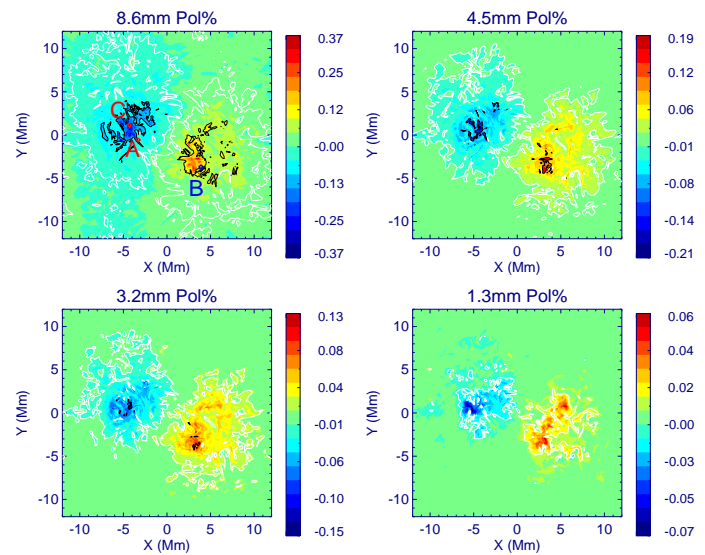}
      \caption{Maps of simulated circular polarization degree for the same wavelengths as in Fig~\ref{fig1}. White and black contours indicate the polarization degrees $|P|=0.01$\% and $|P|=0.1$\%, respectively. Three locations analyzed in Figs.~\ref{fig_tb}, \ref{fig_p}, and \ref{fig_mf} are labeled with the letters ''A'', ''B'' and ''C'' in upper left panel. The color bars associated with the individual panels cover different ranges.
              }
         \label{fig2}
   \end{figure}

      \begin{figure*}
  \centering
            \includegraphics[width=0.22\textwidth,angle=90]{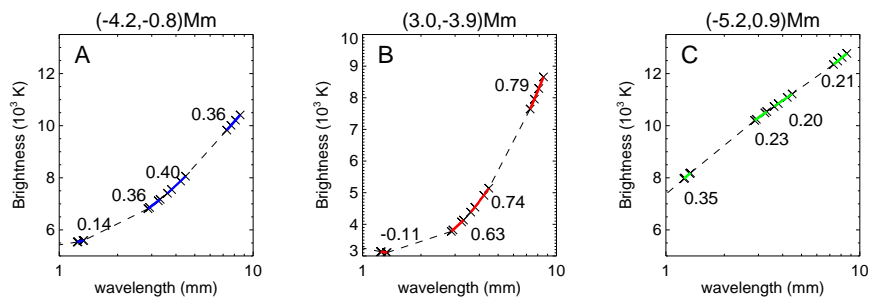}
      \caption{Simulated total brightness spectrum at mm wavelengths for spatial locations (A) (-4.2,0.8)~Mm, (B) (3.0,-3.9)~Mm, and (C) (-5.2,0.9)~Mm. The locations are labeled with the letter ''A'', ''B'', or ''C'' in Fig.~\ref{fig2}. Crosses show the wavelengths of the 4 ALMA bands analyzed in detail. The values of the local slope $n$ are given for each band.
              }
         \label{fig_tb}
   \end{figure*}

 \begin{figure*}
  \centering
            \includegraphics[width=0.22\textwidth,angle=90]{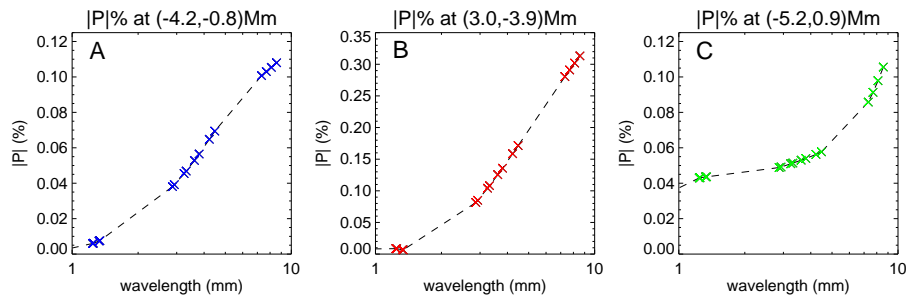}
      \caption{Absolute value of simulated circular polarization degree as a function of wavelength for the same spatial locations as in Fig.~\ref{fig_tb}.
              }
         \label{fig_p}
   \end{figure*}

 \begin{figure*}
  \centering
            \includegraphics[width=0.22\textwidth,angle=90]{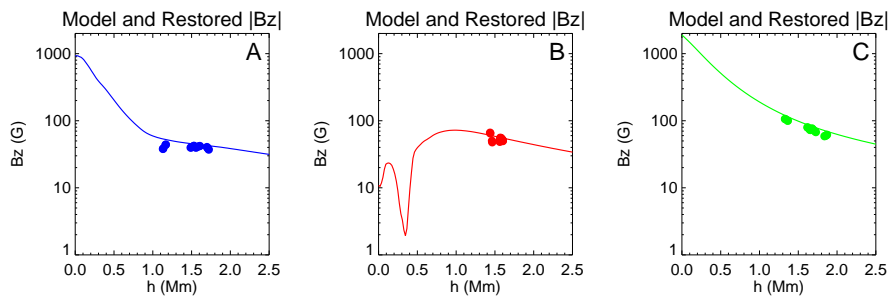}
      \caption{Absolute value of the model longitudinal magnetic field as a function of height (curves). Circles indicate the magnetic field derived from the mm brightness spectrum and polarization degree plotted at the effective formation heights of the mm radiation. The spatial locations are the same as in Figs.~\ref{fig_tb} and \ref{fig_p}.
      }
         \label{fig_mf}
    \end{figure*}

\subsection{Total brightness and degree of polarization}

In Figs.~\ref{fig1} and \ref{fig2} we show the results of the calculations for the total brightness and circular polarization degree at
8.6~mm (ALMA band 1), 4.5~mm (band 2),  3.2~mm (band 3), and 1.3 mm (band 6). As can be seen from Fig.~\ref{fig1} and Table~\ref{table:1} mm/submm radiation samples different temperature regimes producing average brightness in the range from $<T_b>\simeq4500$~K at 0.47~mm up to $\sim8200$~K at 8.6~mm. The corresponding heights to which the sampled thermal structure refers range from about $750$~km to $2050$~km (see last column of Table~\ref{table:1}). As seen from Table~\ref{table:11}, for the analyzed 3D MHD snapshot, which represents quiet-Sun enhanced network, the polarization degree on the absolute temperature scale does not reach $0.5$\% even at the longest wavelength of 8.6~mm. However, the corresponding degrees of polarization seen by the interferometer (see Sect.~\ref{alma}) should range from about 1\% at 1.3 mm wavelength to $\sim5$\% at $\lambda\,=\,$8.6~mm. Maximum polarization is reached in regions of enhanced brightness $T_b$ (corresponding to steep temperature gradients) and of strong magnetic field (causing stronger separation of the effective emitting levels of the two natural modes). In Fig.~\ref{fig2} the white and black contours outline the levels of (absolute) polarization $|P|=0.01$\% and $|P|=0.1$\%, respectively. The maximum of polarization degree and relative numbers of points satisfying the above criteria are listed in Table~\ref{table:11}. At 8.6~mm (upper left panel in Fig.~\ref{fig2}) the degree of polarization exceeds $0.01$\% in 54\% of the points, but exceeds $0.1$\% in only 4\% of the points. At 1.3~mm (lower right panel in Fig.~\ref{fig2}) $P\geq0.01$\% is only found in 9\% of the points. At 1.3~mm and lower wavelengths the maximum degree of polarization (on the absolute scale) does not reach $0.1$\%.



\subsection{Magnetic field from free-free emission}

We employed two different methods to derive the local spectral index $n$ required for $B_l$ calculations. In the first method, the value of the spectral index $n$ for each wavelength was calculated according to \citet{bogod80} by taking the mean value of the two spectrum slopes, which were found using the longer and shorter neighboring wavelengths following Eq.~\ref{n}. In the second method, the spectral index was obtained from a power-law fit of the local brightness spectrum for each of the four frequency bands being considered. Both methods provide similar spectral indices. First we discuss the results of $B_l$ calculations obtained for three selected individual locations, which are labeled in Fig.~\ref{fig2} with letters ''A'', ''B'', and ''C'', and later we consider all spatial locations. For the purpose of this exercise, we assume that the interferometer measurements of circularly polarized flux can successfully be converted to the absolute brightness temperature scale using the single-dish data with sufficient fidelity to measure the spectral indices.

The results of the local spectrum fit (i.e., the second method) for the three locations indicated in Fig.~\ref{fig2} are indicated by (colored) solid lines in Fig.~\ref{fig_tb} together with the simulated $T_b$ values (crosses). The values of the spectral index $n$ are also given in the frames of Fig.~\ref{fig_tb}. Figure~\ref{fig_p} shows the corresponding circular polarization degree $P$ as a function of wavelength for the same spatial locations. The derived values of $n$ and $P$ are listed in Table~\ref{table2} together with the values of the restored field $B_l^{ff}$ and model field at the photospheric level $B_l^{mod}(0)$. As the brightness spectrum gets steeper with wavelength (see Fig.~\ref{fig_tb}), the spectral index $n$ grows (e.g., for the location labeled ''B'' from $n=-0.11$ at 1.3~mm to $n=0.79$ at 8.6~mm), but does not reach $1.0$ in the mm wavelength range (see Table~\ref{table2}).

In Fig.~\ref{fig_mf} the circles denoting the restored $B_l^{ff}$ (referring to the effective formation heights in the 4 ALMA bands) are overlaid on the model profile of $B_l^{mod}$ for the same three locations as indicated in Fig.~\ref{fig2} and analyzed in Figs.~\ref{fig_tb} and \ref{fig_p}. For the displayed locations the restored $B_l^{ff}$ values lie slightly lower than the $B_l$ taken from the model. The normalized difference between $B_l^{ff}$ and $B_l^{mod}$ does not exceed 10-15\% (depending on wavelength) of the $B_l^{mod}$ value with the exception of the location labeled ''A'', where $B_l^{ff}$ obtained from the shortest wavelengths display discrepancies of about 25\%.

In Fig.~\ref{fig3} we show the results for all spatial locations in the form of images of the longitudinal component of the magnetic field obtained from Eq.~\ref{P_n} for the four wavelengths, employing $n$ derived from the second method; we discuss these results further in Sect.~\ref{sect5.3}.

\begin{table*}[!htb]
\caption{Simulated circular polarization, $P$, spectral index $n$, and restored magnetic field, $B_l^{ff}$, in 4 ALMA frequency bands for the 3 lines of sight indicated in Fig~\ref{fig2}. The photospheric value of the field, $B_l^{mod}(0)$, is given for each location. \label{table2}}
\centering
\begin{tabular}[!htb]{ll|ccc|ccc|ccc}
\hline \hline
\multicolumn{2}{c|}{} & \multicolumn{3}{c|}{''A'', $B_l^{mod}(0)=-923$~G} &\multicolumn{3}{c|}{''B'', $B_l^{mod}(0)=11$~G} &\multicolumn{3}{c}{''C'', $B_l^{mod}(0)=-1896$~G}\\
&ALMA bands& $P$,\% & $n$ &$B_l^{ff}$,G &$P$,\% & $n$ &$B_l^{ff}$,G& $P$,\% & $n$ &$B_l^{ff}$,G\\
\hline
\\
\noalign{\smallskip}

\multicolumn{2}{c|}{$8.6$mm, band~1}&  -0.10 & 0.36 & -39 &  0.30 & 0.79 & 51 &  -0.10 & 0.21 & -60\\
\multicolumn{2}{c|}{$4.5$mm, band~2}&   -0.06 & 0.40 & -41     & 0.15 & 0.74 & 53       & -0.06 & 0.20 & -72\\
\multicolumn{2}{c|}{$3.2$mm, band~3}&   -0.04 & 0.36 & -41     & 0.09 & 0.63 & 52        & -0.05 & 0.23 & -76\\
\multicolumn{2}{c|}{$1.3$mm, band~6}&   -0.01 & 0.14 & -41     & -0.01 & -0.11 & 57      & -0.04 & 0.35 & -103 \\
\hline
\end{tabular}
\end{table*}

\begin{table*}
\caption{Maximum absolute value of circular polarization degree (in \%) and maximum absolute value of model $B_l$ in Gauss for a number of analyzed wavelengths as a function of spatial resolution.}             
\label{table3}      
\centering                          
\begin{tabular}{l c c c c |c c c c c c c c}        
\hline\hline                 
\\
Resolution  & \multicolumn{4}{c|}{$|P|$, \%} & \multicolumn{4}{c}{$|B_l|$, G}\\  
 (arc sec) & 8.6~mm & 4.5~mm & 3.2~mm & 1.3~mm & 8.6~mm & 4.5~mm & 3.2~mm & 1.3~mm\\
 \\
\hline                        
\\
\noalign{\smallskip}

model (0.12) & 0.38 & 0.21 & 0.15 & 0.07 & 85 & 91 & 96 & 133\\
0.2 & 0.30 & 0.18 & 0.13 & 0.07 & 85 & 90 & 95 & 129\\
0.4 & 0.24 & 0.15 & 0.11 & 0.06 & 84 & 89 & 93 & 124\\
1.0 & 0.21 & 0.11 & 0.08 & 0.04 & 78 & 83 & 86 & 113\\
4.0 & 0.12 & 0.06 & 0.05 & 0.02 & 48 & 50 & 52 & 59\\
\noalign{\smallskip}
\hline                                   
\end{tabular}
\end{table*}

\subsection{Comparison of restored and model magnetic field}\label{sect5.3}

\begin{figure}
  \centering
            \includegraphics[width=0.4\textwidth,angle=90]{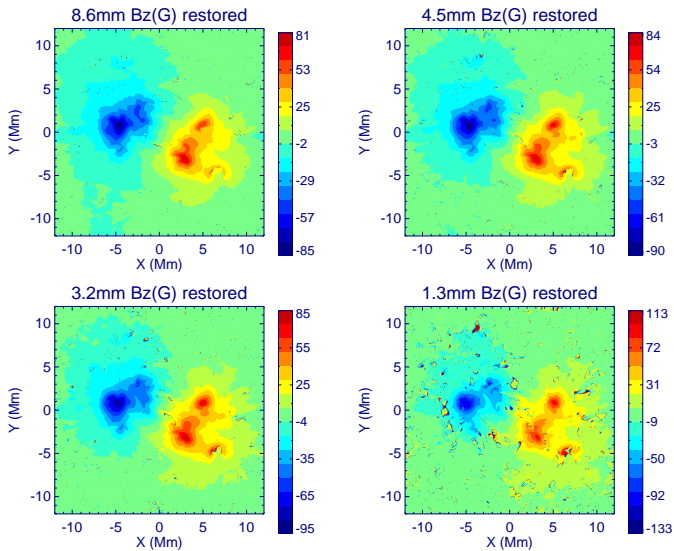}
      \caption{Longitudinal component of the magnetic field, $B_l$,  restored from the synthetic brightness temperature and polarization maps, i.e., $B_l^{ff}$. The wavelengths are the same as in Fig.~\ref{fig1}.
              }
         \label{fig3}
   \end{figure}

\begin{figure}
  \centering
            \includegraphics[width=0.4\textwidth,angle=90]{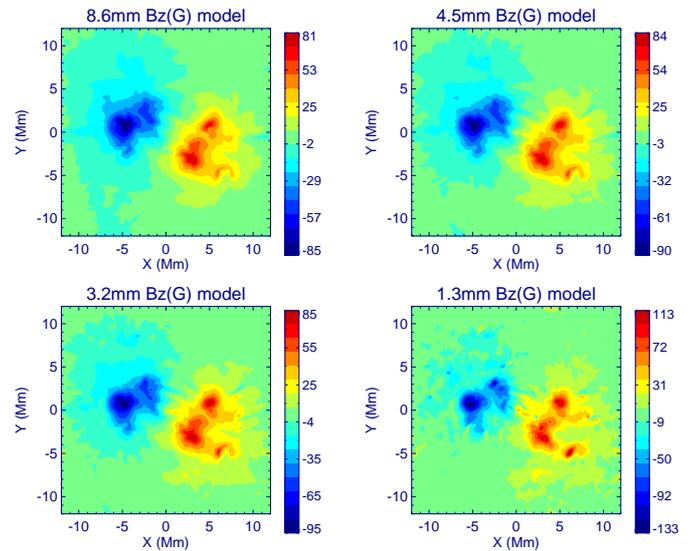}
      \caption{Same as in Fig.\ref{fig3} but with $B_l$ taken directly from the model at the effective heights of the formation of the radiation at the corresponding wavelength, i.e., $B_l^{mod}$.
              }
         \label{fig4}
   \end{figure}

For comparison, the longitudinal component of the magnetic field taken from the model cube at the effective height of formation of each of the four wavelengths being considered is plotted in Fig.~\ref{fig4}. This height is determined on a pixel-to-pixel basis, so that qualitatively the maps from both Fig.~\ref{fig3} and \ref{fig4}, which are shown with the same display range for a given wavelength, look similar, apart from the presence of salt and pepper noise in the simulated magnetic field maps, especially at shorter wavelengths. This noise is due to failure of the algorithm in regions of either low polarization signal or very low spectral index (flat spectrum, see Eq.~\ref{Bl}). Based on the simulated polarization values (see Fig.~\ref{fig2} and Table~\ref{table:1} for details) we employed an arbitrary value of $0.01$\% for the (absolute) polarization signal threshold and considered for further analysis only those locations where the signal exceeds this threshold. In the absence of final circular polarization commissioning data for ALMA we do not know the level of realistic uncertainties for the observational data in this analysis yet.

\begin{figure}
  \centering
         \includegraphics[width=0.35\textwidth]{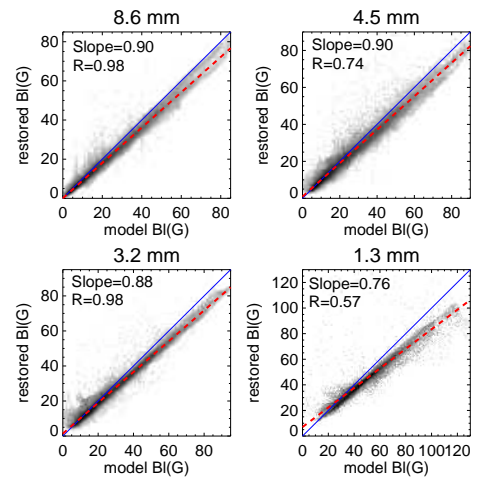}
      \caption{Density scatter plots (2D histograms) of restored longitudinal magnetic field $B_l^{ff}$ vs. model longitudinal magnetic field $B_l^{mod}$ taken at the effective formation heights for the wavelengths 8.6, 4.5, 3.2, and 1.3~mm. The wavelength is indicated at the top of each panel. Darker shading indicates a higher density of pixels in the bin. Solid lines denote the expectation value, $B_l^{ff}=B_l^{mod}$. Dashed lines depict the linear regressions with the slopes and the Pearson correlation coefficients given in the upper left corner of each frame.
}
         \label{fig_scatter}
   \end{figure}

\begin{figure}
  \centering
            \includegraphics[width=0.4\textwidth,angle=90]{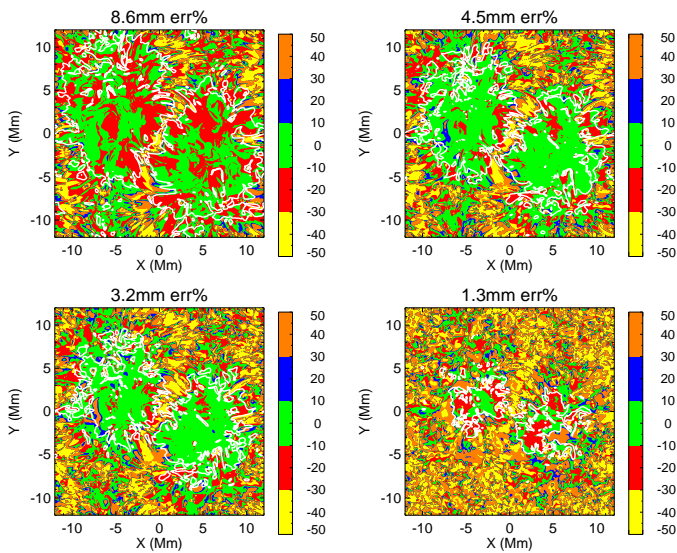}
      \caption{Percent error in the restored longitudinal magnetic field at four wavelengths. White contours denote $P=0.01$\%.
              }
         \label{fig_err}
   \end{figure}

In general, there is a very good correlation between the derived field, $B_l^{ff}$, and the model field, $B_l^{mod}$, as can be judged from Fig.~\ref{fig_scatter}, where we correlate $B_l^{ff}$ with $B_l^{mod}$ taken at the heights corresponding to the effective
formation heights at wavelengths 8.6~mm, 4.5~mm, 3.2~mm, and 1.3~mm. To avoid saturation in the images owing to the large number of pixels in the snapshot, we
show the correlations by binning the data into 2D histograms.
At all wavelengths analyzed, the $B_l^{ff}$ values are systematically lower than the model values $B_l^{mod}$ taken at the corresponding effective heights. The linear slopes derived from linear regression analysis are found to be less than unity and range from 0.90 at 8.6~mm to 0.76 at 1.3~mm (see Fig.~\ref{fig_scatter}). One potential explanation for this effect is that the derived effective heights, from which the radiation is considered to be emitted, are lower than the actual formation heights. This assumption will be further tested in the next paper of the series together with the time-series analysis of the simulated mm/submm brightness. The correlation between the derived and model magnetic field is more tightly constrained for longer wavelengths. The corresponding Pearson correlation coefficient is very close to unity (0.98) at 8.6~mm and 3.2~mm. At 4.5~mm the distribution slightly widens producing a correlation coefficient of 0.74, while at 1.3~mm a large number of outliers in the derived $B_l^{ff}$ values result in the rather weak correlation coefficient of 0.57.

\begin{figure*}
  \centering
            \includegraphics[width=0.3\textwidth,angle=90]{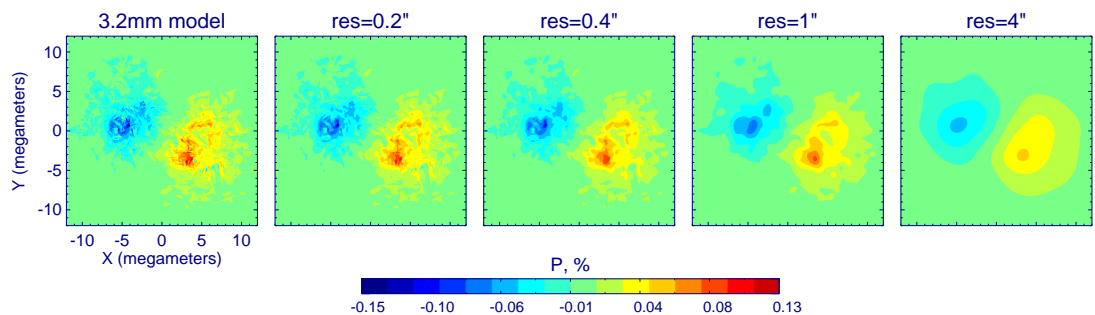}
      \caption{Simulated degree of circular polarization for 3.2~mm at the resolution of the model and at resolutions of 0.2\arcsec, 0.4\arcsec, 1\arcsec, and 4\arcsec.  }         \label{fig_pol_conv}
   \end{figure*}

 \begin{figure*}
  \centering
            \includegraphics[width=0.3\textwidth,angle=90]{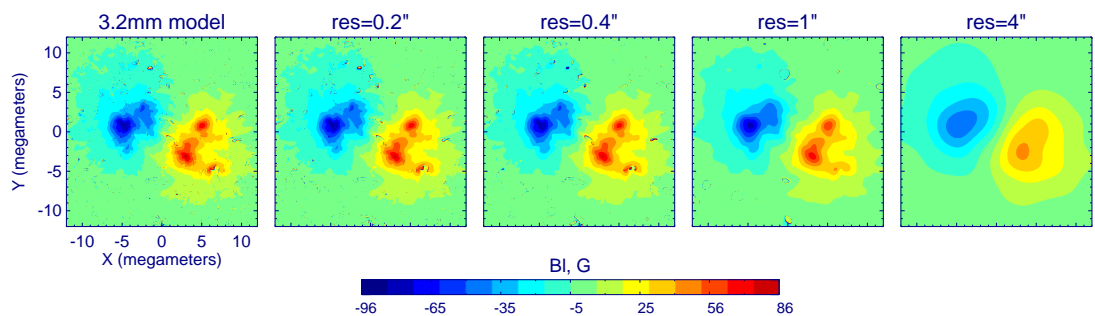}
      \caption{Longitudinal magnetic field restored from 3.2~mm simulated polarized emission at the resolution of the model and at resolutions of 0.2\arcsec, 0.4\arcsec, 1\arcsec, and 4\arcsec.}
         \label{fig_mf_conv}
   \end{figure*}

\begin{figure}
  \centering
         \includegraphics[width=0.35\textwidth]{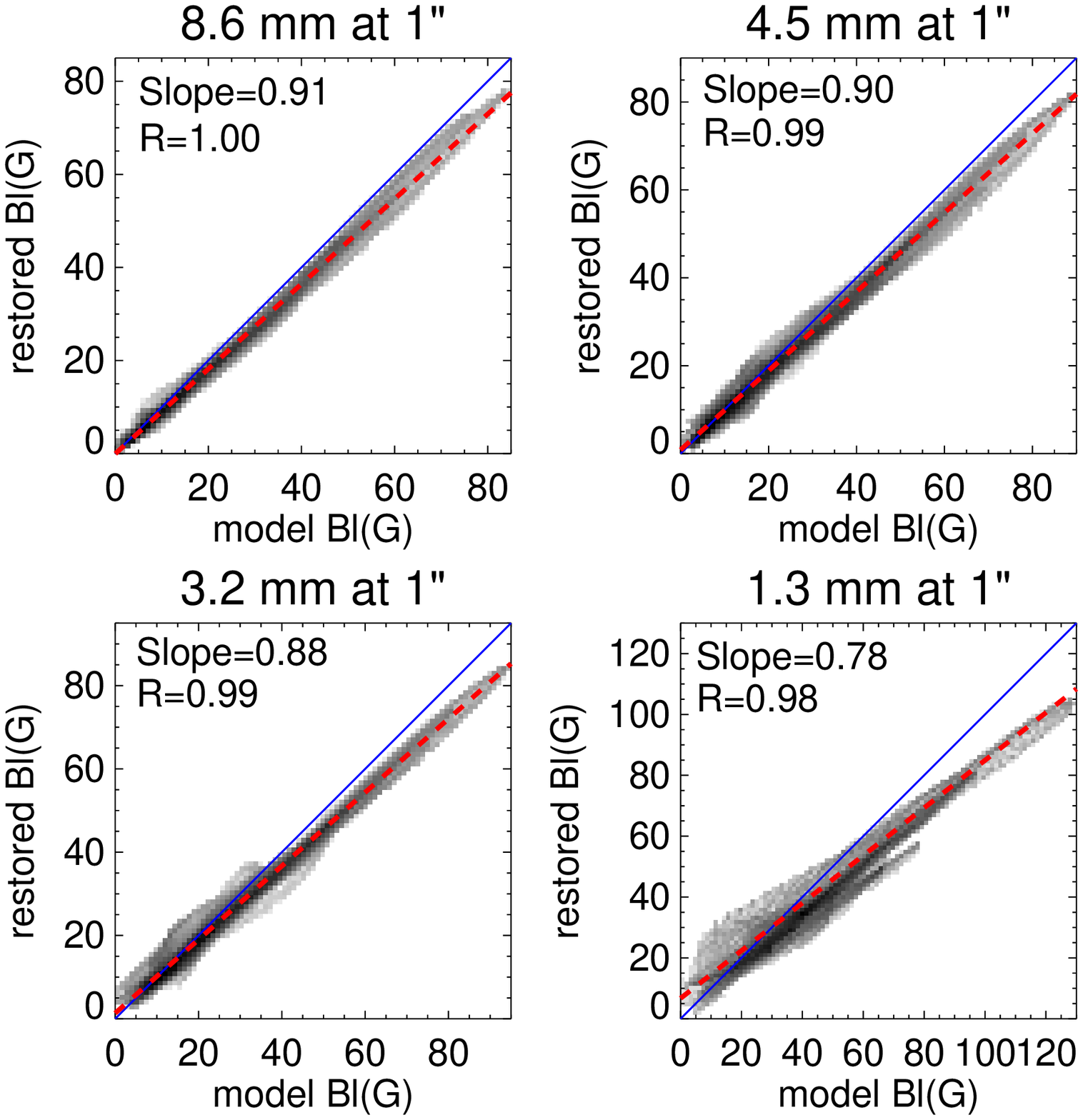}
      \caption{Density scatter plots (2D histograms) of the longitudinal magnetic field, $B_l^{ff}$, restored from the simulated data at a resolution of 1\arcsec\ vs. the model longitudinal magnetic field, $B_l^{mod}$, smeared to 1\arcsec\ and sampled at the effective formation heights for the wavelengths 8.6, 4.5, 3.2, and 1.3~mm (as indicated above the individual panels). Darker shading indicates a higher density of pixels in the bin. Solid lines denote the expectation value, $B_l^{ff}=B_l^{mod}$. Dashed lines depict the linear regressions with the slopes and the Pearson correlation coefficients given in the upper left corner of each frame. 
              }
         \label{fig_scatter_1arcsec}
   \end{figure}

We calculate the percent error distribution, which we introduce as $err=(B_l^{ff}-B_l^{mod})/B_l^{mod}$ for every location, to provide a 2D quantitative comparison; we depict this 2D comparison in Fig.~\ref{fig_err}. Each color represents a certain error range, as
indicated by the color bar. The white contours represent locations with $|P|=0.01$\%. Within the regions with $|P|>0.01$\%, $B_l^{ff}$, and $B_l^{mod}$
agree reasonably well with the relative error not exceeding $10$\% in the majority of points at all wavelengths. There is a tendency for the $B_l$ values to be underestimated within these contours. This is most clearly seen at 8.6~mm (top left panel in Fig.~\ref{fig_err}), where 51\% of points within the region with $|P|>0.01$\%  have negative $err$ exceeding 10\% (shown with red color).

\subsection{Effect of spatial resolution}

The spatial resolution of images produced by an interferometric array such as ALMA depends on the array configuration (the number of antennae and maximum baseline) and wavelength. Observations by ALMA can achieve spatial resolutions in the range 0.005\arcsec\ - 5\arcsec\ , although the highest resolution may be difficult to use in practice for solar data.
To account for this finite resolution in a simple fashion, we spatially smeared the synthetic brightness and polarization maps by convolving with a Gaussian kernel of corresponding FWHM to mimic the instrumental profile. The effect of spatial smearing on the circular polarization and derived $B_l^{ff}$ is illustrated in Figs.~\ref{fig_pol_conv} and \ref{fig_mf_conv}, respectively, for 3.2~mm and for the following four values of FWHM: 0.2\arcsec, 0.4\arcsec, 1\arcsec\ and 4\arcsec. Whereas at 0.2\arcsec\ much of the fine structure present in the polarization and magnetic field of the original-resolution images is preserved, it is increasingly lost as the resolution is reduced further. At the same time the polarization signal is rapidly reduced, as can be seen from Table~\ref{table3}, where we list the maximum degree of the polarization and maximum values of the model $B_l$ at the resolution of the model and as a function of the spatial resolution.
There is a significant change when going to 1\arcsec\ resolution. Thus, at 3.2~mm the polarization goes down to almost 50\% of the original signal and most of the fine structure is already smeared out, whereas the larger scale pattern is readily discernible and closely reflects that in the original image. The sampled range of $B_l$ values is less affected by the decreased resolution, going down only slightly to 90\% of the original maximum $B_l$ signal at a resolution of 1\arcsec. Finally, at 4\arcsec\ resolution the polarization signal at 3.2~mm drops dramatically to 30\% of the original value and the maximum of $B_l$ drops to roughly half of the original value. The effect of spatial smearing is expected to be similar for both the degree of polarization calculated on the absolute temperature scale and for the degree of polarization measured by the interferometer.

   \begin{figure*}
  \centering
            \includegraphics[width=0.22\textwidth,angle=90]{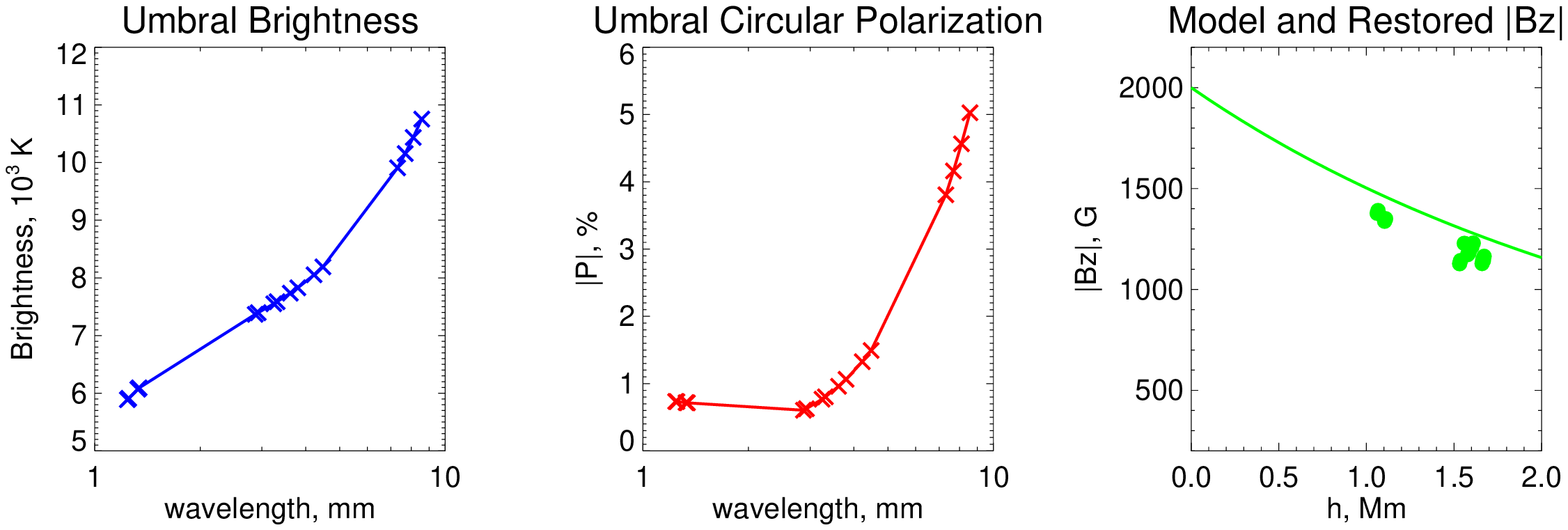}
      \caption{Left and center: total brightness spectrum and degree of circular polarization calculated from the umbral model by \citet{Severino} at mm wavelengths. Crosses show the wavelengths of the 4 ALMA bands. Right:  absolute value of the model longitudinal magnetic field as a function of height. Circles indicate the derived magnetic field plotted at the effective formation heights of the mm radiation.
              }
         \label{fig_sev}
   \end{figure*}

The quality of the $B_l$ restoration is not significantly influenced by the spatial smearing as is demonstrated in Fig.~\ref{fig_scatter_1arcsec}, where we show density scatter plots of the longitudinal magnetic field $B_l^{ff}$ restored from the simulated data at a resolution of 1\arcsec\ versus the model longitudinal magnetic field $B_l^{mod}$ smeared to 1\arcsec\ and taken at the effective formation heights of the wavelengths 8.6, 4.5, 3.2, and 1.3~mm. The field restored from smeared data shows less scatter than the field from the original data. Consequently, the Pearson correlation coefficient derived from smeared data is higher than the coefficient from the original data at all wavelengths investigated. Extreme $B_l$ values are smoothed out, but the linear slopes remain almost identical to those of the original data. On the whole, the overall relation is largely unchanged and the diagnostic remains robust.

\section{Discussion and conclusions}\label{discuss}

We studied the circularly polarized emission expected in chromospheric observations at mm/submm wavelengths using realistic 3D simulations of the quiet Sun chromosphere in an enhanced
network region. We have demonstrated how the LOS chromospheric magnetic field can be recovered from circular polarization observations. We investigated how reliable this technique is for the recovery of the simulation magnetic field and studied the effect of instrumental resolution on the results. The computed circular polarization is rather small on the absolute temperature scale, but more reasonable for interferometer observations, which are not sensitive to the large-scale disk component, ranging from about 1\% at 1.3 mm to 5\% at 8.6 mm.  If the results of this simulation are valid for the solar atmosphere, then sensitivity at better than 0.1\% in degree of circular polarization may be required for robust application of this technique. Larger differences in the height of formation of the two natural modes will result in larger degrees of circular polarization than we infer here. The effect of instrumental smearing reduces polarization further, leading to values less than $1$\% for 3~mm interferometer observations at 1\arcsec\ resolution and even lower polarization at the shorter wavelengths. This suggests the importance of the longer millimeter-wavelength observations ($>$ 4 mm) for reliable polarization measurements in the QS and, consequently, for tomography of the magnetic field at chromospheric heights.

Higher values of circular polarization have been observed at longer mm wavelengths in ARs. Thus, \citet{kundu} detected polarized emission in an AR at $\lambda\,=\,$9~mm with $P$ in the range from about 1 to 4\%. We have investigated the expected circular polarization from active regions at millimeter wavelengths using two sets of umbral models:  the umbral model by \citet{Severino}, which provides the best agreement with the observed mm brightness in umbrae according to \citet{LSC14}, and a more recent semiempirical umbral model by \citet{Fontenla2011}. The magnetic field was modeled by a vertical dipole buried under the photosphere, following \citet{zlotnik}. The results of the millimeter polarized emission calculations for the umbral model by \citet{Severino}, together with the comparison of the model and restored longitudinal magnetic field, are presented in Fig.~\ref{fig_sev}. The simulated polarization in the AR is higher than typical values in the QS. This is explained by higher $B$ values and by steeper gradients of temperature in umbral atmosphere compared to the QS, resulting in a steeper brightness temperature spectrum. For a sunspot with photospheric magnetic field of 2000~Gauss and a dipole depth of 10000~km, the umbral model by \citet{Severino} predicts a circular polarization degree of 5\% at 9~mm and of 1\% at around 3~mm (Fig.~\ref{fig_sev}), while the model by \citet{Fontenla2011} results in 6\% and 3\%, respectively. These results are in line with the findings of \citet{fleishman2015}, who simulated umbral radio emission from AR 12158 at millimeter wavelengths using an extrapolated photospheric magnetic field and obtained a degree of circular polarization of $3.6$\% at 3~mm. As was true in the QS analysis here, the umbral longitudinal magnetic field $B_l^{ff}$ derived from the mm data is found to be about 10\% lower than the model field $B_l$ (see right panel of Fig.~\ref{fig_sev}).

The method relies on high--resolution and high--precision circular--polarization observations, which can be provided by millimeter interferometers with linearly polarized feeds like ALMA. The interpretation of the polarization measurements in terms of chromospheric magnetic fields will require excellent knowledge of the instrumental polarization characteristics, including the beam responses of both polarizations and polarization leakage terms.

Our test of the method of probing the longitudinal chromospheric magnetic field using free-free emission at mm/submm wavelengths has found that (albeit without realistic instrumental uncertainties) the method is applicable to measurement of the weak QS (and stronger AR) magnetic field in the chromosphere, recovering the longitudinal component of the field at the effective height of formation.  The restored field is averaged over the heights contributing to the radiation weighted by the strength of the contribution. This results in relative errors generally below 10\% in regions where (absolute) circular polarization exceeds 0.01\%. However, the restored values of $B_l^{ff}$ are on average systematically lower than the model $B_l$ values taken at the effective formation heights.

This systematic discrepancy will be further investigated in a separate study that will seek to understand and address its cause. Our results indicate that the effect of instrumental smearing does not significantly influence the quality of restoration of the longitudinal field except that it smooths out extreme values. The method holds the potential to determine fields in QS and AR at chromospheric heights. It provides diagnostics of the 3D structure of the longitudinal component of the magnetic field in the chromosphere, when applied to multiple mm wavelengths, or when combined with magnetic field measurements using atomic lines.

\begin{acknowledgements}
We thank Andreas Lagg for computational support. The research leading to these results has received funding from the European Research Council under the European Union's Seventh Framework Programme (FP7/2007-2013) / ERC grant agreement nr. 291058 and from the Research Council of Norway through the project "Solar Atmospheric Modelling" and a grant of computing time from the Program for Supercomputing. This work was partly supported by the BK21 plus program through the National Research Foundation (NRF) funded by the Ministry of Education of Korea, NSF grant AGS-1250374,  NSF grant AST-1312802, NASA grant NNX14AK66G, and by Russian RFBR grants 15-02-03835 and 16-02-00749. M.~Loukitcheva acknowledges Saint-Petersburg State University for research grants 6.42.1428.2015 and 6.37.343.2015.

\end{acknowledgements}

\end{document}